# Inverting the SerDes Link Design Flow Process

Michael J. Degerstrom, Mayo Clinic
degerstrom.michael@mayo.edu

Chad M. Smutzer, Mayo Clinic
smutzer.chad@mayo.edu

Patrick J. Zabinski, Mayo Clinic
zabinski.patrick@mayo.edu

Barry K. Gilbert, Mayo Clinic
gilbert.barry@mayo.edu, 507-284-4056

## Abstract

The traditional SerDes link simulation process begins with the extraction of printed circuit board (PCB) physical stripline and via models, followed by channel modeling and link simulation. We invert this simulation flow by first creating "link performance curves" across an array of hypothetical channels defined with specially-developed, high level, equation-based models; limited physical extraction is later undertaken to relate PCB channel implementation to these performance curves. These curves allow us to determine the system-level SerDes channel requirements and to become better informed in choosing PCB technologies for lower cost and easier manufacturability. The inverted modeling process is very efficient, allowing for the rapid identification and avoidance of problematic channel topologies and the study of other potentially useful channel designs.

## Biographies

**Michael J. Degerstrom** received a BSEE from the University of Minnesota. Mike is currently a Senior Engineer at the Mayo Clinic Special Purpose Processor Development Group. His primary area of research and design has been in the specialty of signal and power integrity.

**Chad M. Smutzer** received a BSEE from the University of Iowa in Iowa City. He is currently a Senior Engineer at the Mayo Clinic Special Purpose Processor Development Group where he performs signal and power integrity analysis.

**Patrick J. Zabinski** received a BSEE from St. Cloud State University (St. Cloud, MN), MSEE from University of North Dakota (Grand Forks, ND), and MBA from Minnesota School of Business (Richfield, MN). He is a principal engineer at the Mayo Clinic Special Purpose Processor Development Group, where he leads research initiatives that advance and integrate novel electronics technologies.

**Barry K. Gilbert** received a BSEE from Purdue University (West Lafayette, IN) and a Ph.D. in physiology and biophysics from the University of Minnesota (Minneapolis, MN). He is currently Director of the Special Purpose Processor Development Group, directing research efforts in high performance electronics and related areas.

## Introduction

Signal integrity (SI) engineers of today's highest serial bandwidth systems are faced with challenges such as enabling >30 Gb/s non-return-to-zero (NRZ) and 28 Gb/s pulse amplitude modulation (PAM)-4 links. Moreover, given the significant demand for digital bandwidth, the number of links per system continues to grow, which in turn increases printed circuit board (PCB) routing congestion and crosstalk. To address these trends we believe that there is significant advantage in meeting the requisite design goals by first obtaining a broad understanding of serializer/de-serializer (SerDes) capabilities over a variety of potential channel configurations.

Meeting design goals is a multi-faceted endeavor. Of prime importance is that the as-designed SerDes links work at the targeted data rate, which is typically the maximum rate that the SerDes circuits are capable, over "reasonable" channels. Yet another aspect of achieving the design goals is to adapt a PCB stack-up configuration that is easily manufacturable. Typically the PCB fabrication cost is minimal compared to the aggregate cost of the devices supported by the PCB, i.e., the "bill-of-materials."

Therefore we generally do not emphasize the requirement that the PCB be easy to fabricate merely to minimize PCB cost. Instead, we minimize board complexity primarily as a means to ensure that the PCB can be manufactured rapidly. Over many years we have observed that PCBs have often experienced delivery delays due to production problems associated with a complex board design.

The SerDes link modeling approach described herein is essentially the opposite of our previous link analyses approaches. Hence we informally refer to this new approach as an “inverted” design flow. For our previous design flow, we modeled vias such that the reflected voltage (S11) was less than approximately -12 dB, and then limited the via count per channel to two vias, or four vias if blocking capacitors were used. We would then specify one of the lowest loss materials available that would be conducive to the manufacture of high-layer-count PCBs, followed by simulations of potentially marginal links using extracted models. Only rarely would we attempt higher level link performance modeling since it would be difficult to do so based upon PCB model extractions. However, our prior design flow did not provide us with significant insight into overall link performance. Nonetheless this approach served us well over the last decade in that no SerDes link failures were detected. In hindsight we were obviously over-designing in many cases, but the added expense of higher-technology PCBs would, in the case of link failure, easily offset the very high cost of debugging and redoing PCB design and assembly. With regard to future designs, this prior design process may not extend well, since the PCBs would require advanced technologies to support the increased numbers of links each operating at higher data rates with wide design margins, a combination of boundary constraints that may not be feasible.

This paper primarily discusses the creation of SerDes link performance curves, which we use as the primary guidance in the generation of design rules that guarantee link performance goals. Further, our newer methodology easily supports a broad understanding of point topics, e.g., whether cheaper via technologies will be adequate to achieve targeted link performance.

## Performance Curve Description

To facilitate discussion, as shown in **Figure 1**, example link performance curves are based on a very simplistic channel that uses three stripline segments stitched together with two vias. This particular channel is “intra-PCB,” meaning that it is contained within a single PCB. These new methodologies are expected to be extensible to more complex channels such as those across backplanes, etc., that require more PCB stripline segments as well as multiple connectors. Later we will describe how we model a channel with its constituent parts, but for now, the link performances are generally defined by a line delineating where SerDes links fail; i.e., performance combinations above/below this line are failing/passing links, respectively. A failing link violates the 40 millivolt high eye mask, as shown in the eye diagram of **Figure 2**, which we chose as the pass/fail criteria. We ignore transmitter (TX) jitter and hence there are no eye width violations. For application to a future system design, TX jitter would be included and the link pass/fail criteria would be chosen as provided by the IC vendor, whether an eye mask, bathtub curve, etc.

We began by neglecting common channel impairments such as the SerDes TX and receiver (RX) capacitance and package models. We anticipated, and later observed, that the performance curves shifted downward and to the left when these constraints were added. However, the simplicity of the channels, as modeled, enabled a better understanding of performance trends and simplified the topics described below.

The baseline performance curves presented in **Figure 1** showed trends for the RX having 0, 3, and 12 decision feedback equalizer (DFE) taps. Data were generated using three differing via models having various levels of complexity, with the details of which will be discussed later. Once the system architect selects an IC technology, one set of performance curves would typically be created using the number of DFE taps available for that IC technology, unless additional constraints are established, such as the requirement to limit SerDes power consumption by using fewer taps. For simplicity, the initial performance curves were calculated using no TX equalization and only DFE for the RX equalization. These performance curves were generally created at a 10 Gb/s data rate, but we have demonstrated that these curves are independent of data rate since, as will be explained later, all electrical performance and physical criteria are scaled to the Nyquist frequency, i.e., the frequency equal to one-half the SerDes data rate. Further, we used the Keysight Advanced Design System (ADS) TX and RX models that apparently do not account for differences in frequency, inferred from the observation that the link performance was frequency-independent.

We have defined link performance curves to be functions of via reflected voltage and total stripline attenuation given in units of dB. The term dB is defined consistently throughout this paper; the included figures document that the dB unit is equal to 20 times the log10 of voltage amplitude. These definitions are the most appropriate and likely will appear so to the general SI design community, because any via stub and other artifacts of the via contribute much of the reflected voltages in a PCB channel. Additionally, stripline losses typically contribute to most of the PCB channel insertion loss. By viewing the baseline link performance curves in **Figure 1**, two trends are obvious. First, it is clear that the DFE is effective, as expected, in overcoming stripline loss. Furthermore, the effect of adding DFE taps has diminishing returns in performance, in that using 3 taps allows for 6 dB of additional stripline loss and using 12 taps is required to gain another 6 dB of stripline loss tolerance. Second, it (erroneously) appears that the reflected voltage from the vias significantly degrades the link performance. However, a surprising result can be inferred from **Figure 1** in that the links are still operational despite very large reflected voltages. The various via models used in the channel model are lossless, or nearly so. Therefore, using conservation of power principles, the reflected voltage can be used to compute the insertion loss of each via assuming that there are no radiated emissions. For example, given a reflected voltage of -3 dB then the via insertion loss is 3 dB. Yet the link is still functional, albeit perhaps barely so, with two of these vias in the channel! To prevent the link from failing, the stripline insertion loss must be decreased by about 6 dB – but this loss offsets the sum of the insertion losses of the two vias. This finding indicates that the very high reflected voltage from the vias had little effect on link performance, i.e., the large voltage reflections did not contribute to close the eye and cause the link to fail. Although the details are not presented here, pulse response reflections between the two vias are shorter in duration and have much lower amplitude than that from a non-reactive discontinuity – an effect that allows links with highly-capacitive vias to remain functional.

Subsequent sections will discuss how we created the models that were used for generating link performance curves. Several point studies will be described that were facilitated by our modeling process. Our progress on simplified crosstalk modeling will then be presented. Finally, the topics discussed above will be summarized, and concluding observations presented.

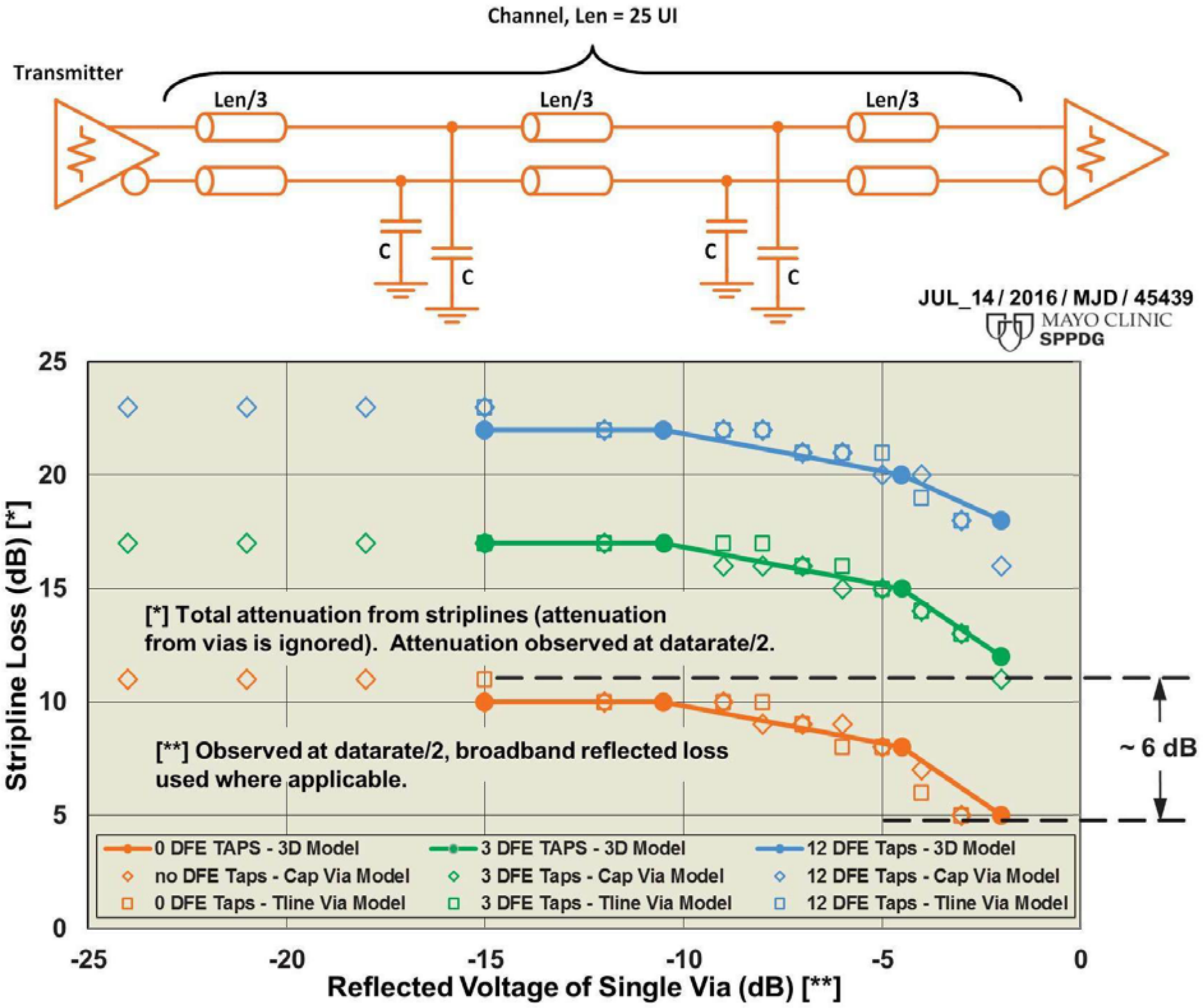

**Figure 1 – Link performance curves for baseline link model [45439]**

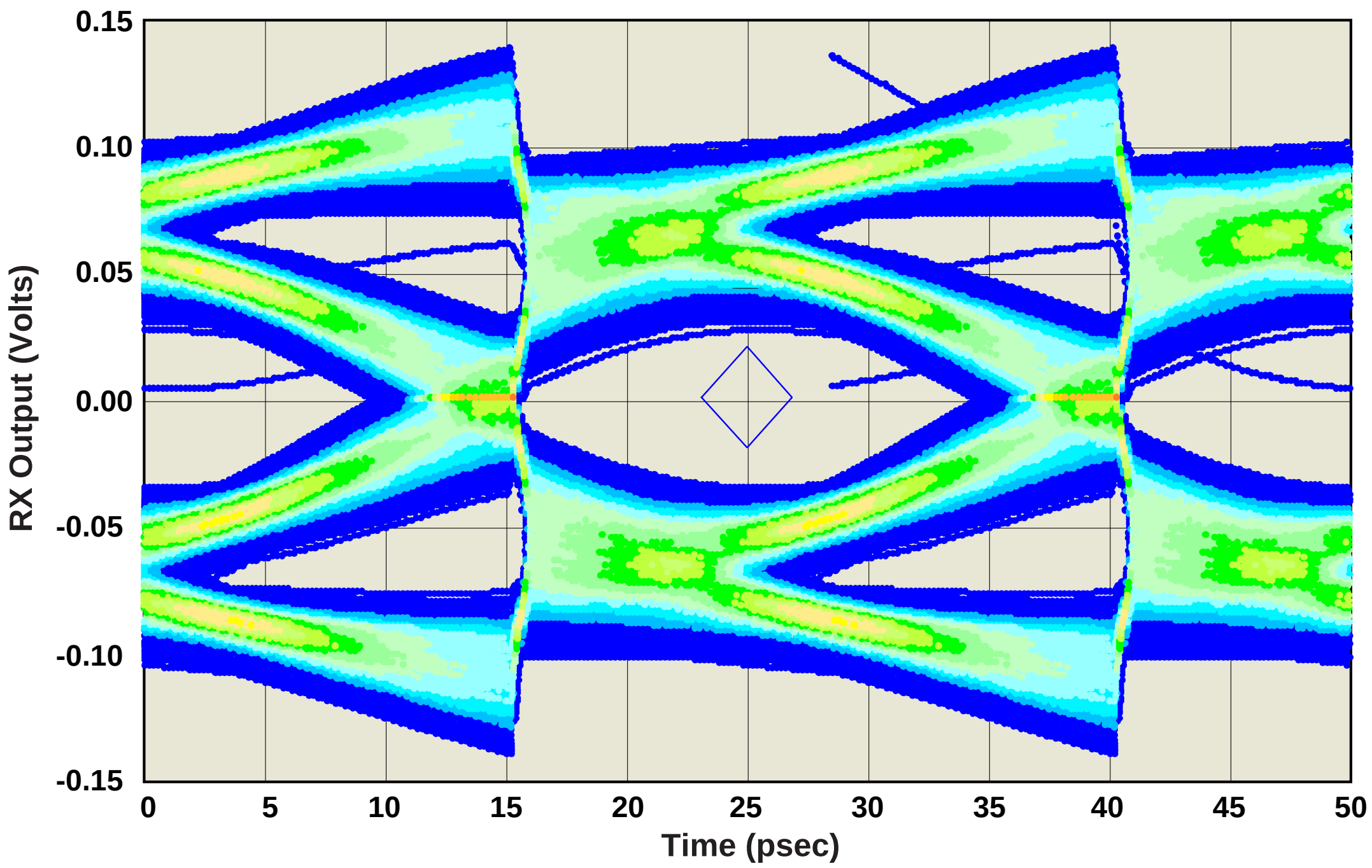

**Figure 2 – Eye diagram with pass/fail mask [45446]**

## SerDes Channel Model Development

The goal of this work was to identify methodologies to simplify the process of generating link performance curves, while simultaneously disassociating link performance from specific PCB material and via technologies. Our intent was to define stripline and via models in the simplest and least-physical terms possible so that the link performance curves could be generalized, thereby achieving several benefits, e.g., making them useful regardless of targeted PCB technologies. Models that are completely disassociated from physical definitions and defined only by electrical performance are termed as being "foundational." Conversely, traditional SI modeling approaches utilize physically-based models almost exclusively for SerDes link analysis.

Our foundational model defined the transmission line in terms of unit interval (UI) length (i.e., a bit period), high-frequency characteristic impedance, and separate dielectric and metal losses. We also specified a frequency, typically the Nyquist frequency, where these losses and electrical length are to be interpreted in terms of frequency-dependent resistance, inductance, capacitance and conductance (RLCG).

The metal losses for the foundational transmission line model were defined as the sum of separate DC and skin-effect (AC) losses. Attempts were also made to add surface roughness loss, but the initial transient responses were not accurate. We acknowledge the need to incorporate surface roughness effects will be critical for high data-rate applications; therefore, we hope to continue improving this model. The transmission line model is defined as a non-coupled differential pair, and is created as a subcircuit in ADS as shown in **Figure 3**-a.

Where:

- f1 = Nyquist frequency (Hz)
- Len = electrical length of  line (UI)
- Lskew = skew between differential lengths (UI)
- Zohf = high frequency characteristic impedance (ohms)
- dBRdc = DC resistive line loss (dB)
- dbRac = AC resistive line loss at f1 (dB)
- dbGac = AC dielectric line loss at f1 (dB)

To account for the dielectric and skin-effect losses, we inverted the equations in **[1]** to solve for distributed RLCG  as a function of frequency.  The resulting transmission line models exhibited expected behavior in both the frequency and transient simulations by obtaining perfect correlation, as shown in **Figure 4**, with the ADS TLINP model, which is defined both in foundational and physical parameters.

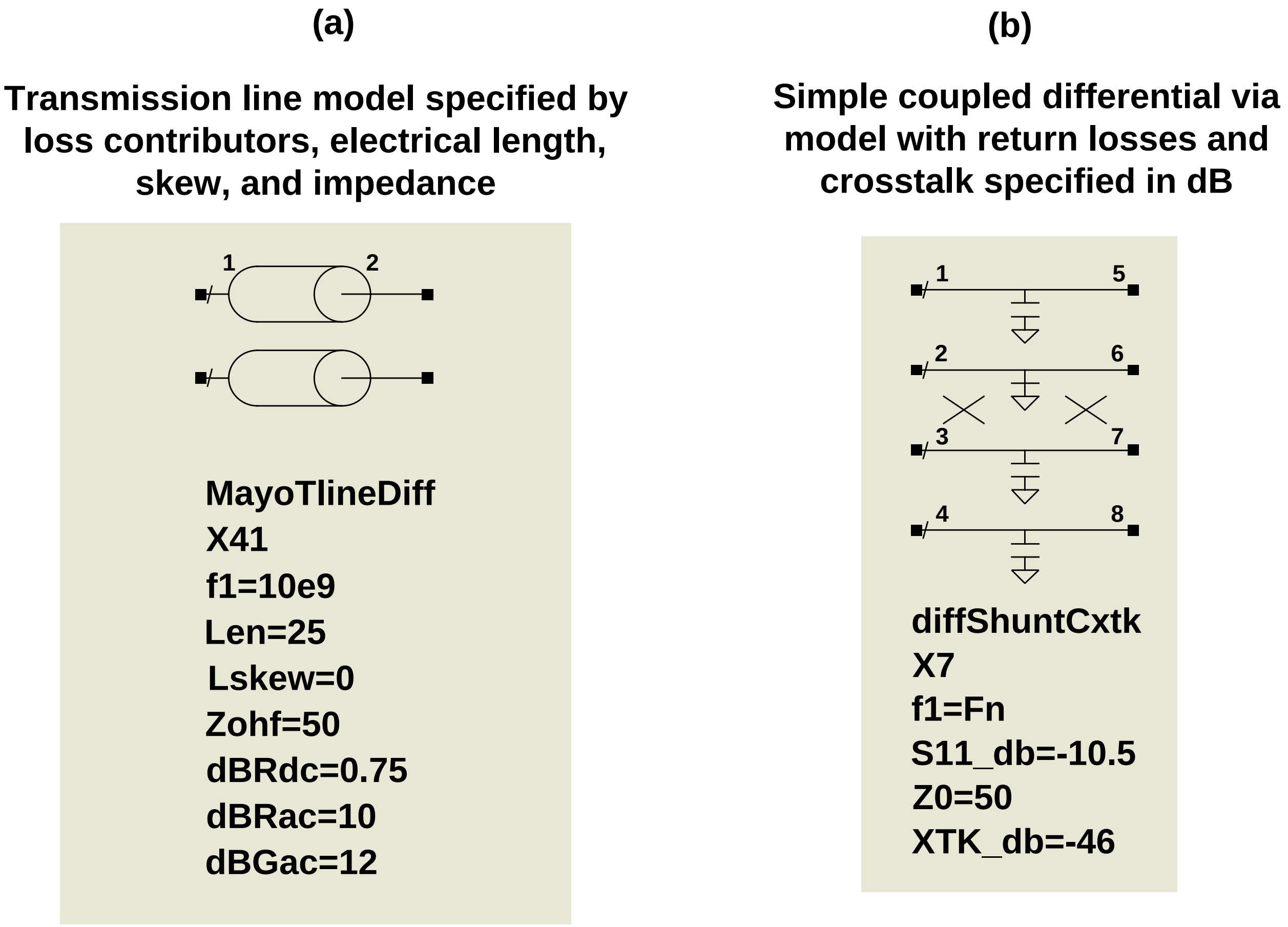

**Figure 3 – Foundational stripline and via models [45442]**

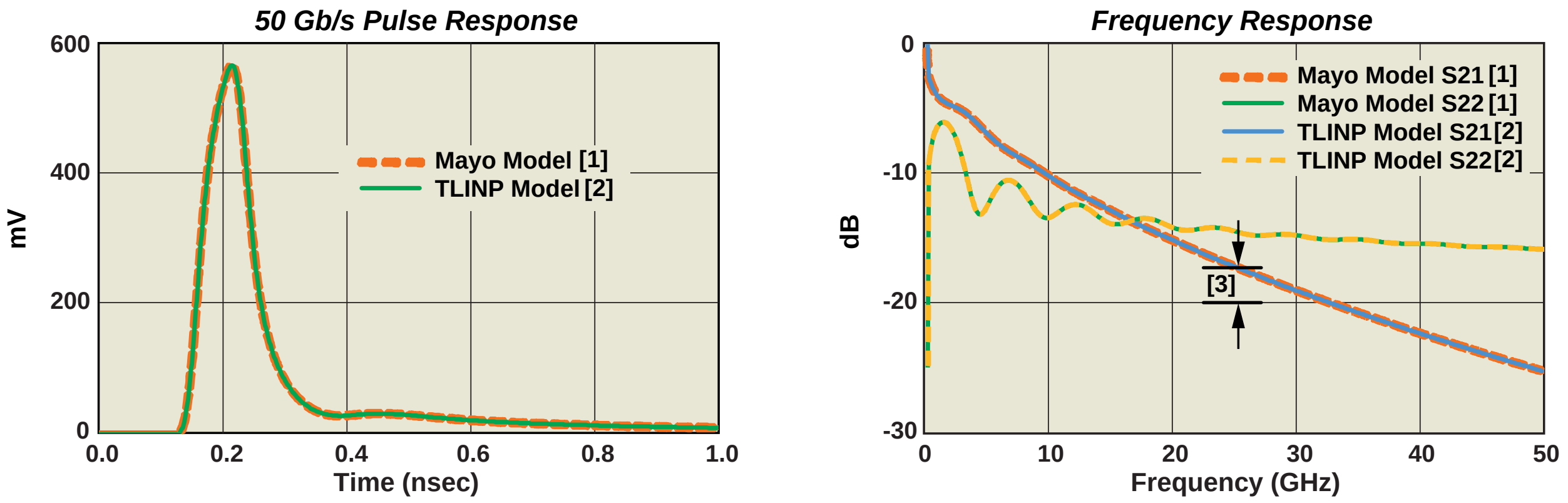

[1] Mayo models set to Rdc = 0 dB, Rac = 20 dB, Gac = 0 dB, Length = 2 UI at 25 GHz
[2] Keysight TLINP models set to match Mayo model settings
[3] Short lossy lines exhibit interaction with 50 ohm port impedance such that insertion losses differ from set values (-17 dB actual versus -20 dB as set in this example)

**Figure 4 – Comparison between foundational stripline and ADS TLINP models [45243]**

Typically, our SI analyses are almost exclusively targeted to SerDes links interconnected with PCB technology. Therefore, critical parts of the SerDes channel include the PCB vias and hence a foundational model was sought for this physical structure. The approach to developing a foundational model for the via was to start with the simplest instantiation and add complexity as needed. The accuracy requirement for the foundational via model is quite austere: if link performance curves obtained using the foundational models match well to those generated with the physical models extracted from full-wave electromagnetic simulation, then the foundational model complexity is assumed to be adequate.

Physical via models described in **[2]** were repurposed to validate the foundational via model. These physical models were based on a specific 26-layer PCB stack-up and different break-out layers and were analyzed to obtain S-parameter models with varying levels of electrical performance. An illustration of this model is depicted in **Figure 5**-a. The model has two differential vias so that crosstalk can also be simulated; however, the added differential vias have minimal effect on via performance. The reflected voltages for four different models are presented in **Figure 5**-b. Overlaid on the reflected voltage curves are straight lines to estimate the broadband reflected voltage (BRV)[1] which is a methodology that we have used to estimate the effective reflected voltage from a dynamic reflected voltage response **[3]**.

[1] In previous publications we have used the term "broadband return loss", or "BRL". We have changed the terminology since reflected voltage is of primary concern for signal integrity analysis whereas return loss may be either a voltage or power term

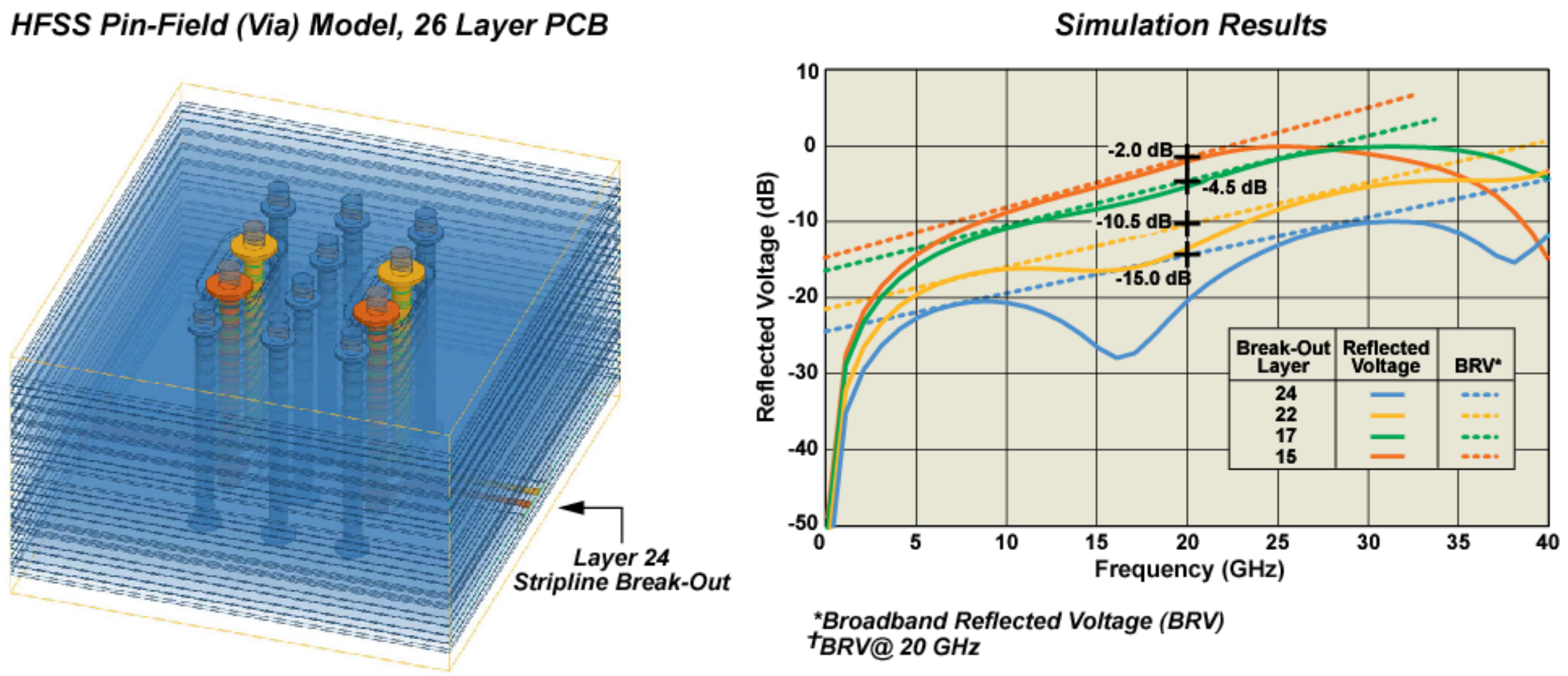

**Figure 5 – Example 3D electromagnetic models to predict reflected voltage (shown) and crosstalk [45450]**

It is well understood that, to a first order, PCB vias with stubs exhibit overly capacitive behaviors. Therefore the simplest circuit representation for the foundational via model is that of a lumped capacitor having a value to match the BRV of the physical via model at a given frequency. The reflected voltage of the lumped capacitor, having a single pole, will exhibit a monotonically increasing voltage magnitude versus frequency and therefore will have little resemblance to the frequency-dependent reflected voltage of the physical via. However, as previously stated, we only care that the foundational and physical models, as implemented into a SerDes channel, result in the same link performance curves.

The ADS implementation of the foundational via model appears in **Figure 3**-b. This model requires the Nyquist frequency and the reflected voltage at this frequency. Simple equations are used to compute analytically the value of the capacitor needed to realize the specified reflected voltage. The model features two identical capacitors of the same value to represent a differential via pair and a repeated capacitor pair, with coupling capacitors connecting these capacitor pairs to provide a means for modeling crosstalk from one aggressor onto a victim via pair.

We created a 2-via channel as presented in **Figure 1** using one of the four physical 3D via models, and then repeated the analysis with the vias implemented as a lumped capacitor model, where the BRV at the Nyquist frequency matched the performance of the 3D representation (while, setting the via crosstalk to a negligible value). We observe, in **Figure 6**, that the channel reflected voltages in the frequency domain correlate between physical and foundational models reasonably well for the two vias with very long stubs, whereas the correlation is quite poor for the vias with the shortest stubs. Conversely, the insertion loss correlation is good for short stubs and poor for the longer via stubs. We also plot 40 Gb/s pulse responses, as shown in **Figure 7**, for that same channel used for the channel loss comparisons. The pulse responses using the physical via models are delayed from those using the capacitor models since the physical via models have some length of stripline egress out of the model, and the via barrel itself adds electrical length compared to that of the lumped capacitor. Generally, the pulse responses resulting from using physical versus capacitive models are similar for the two short via stubs and differ more as the via stub lengthens.

Despite the lack of correlation in both frequency and time domains between passive channel behaviors using physical versus capacitor models, revisiting **Figure 1**, it is apparent that the link performance curves, created with these lumped capacitor models, match well to those produced with the physical via models. It should be noted that the process used to generate these link performance curves is somewhat subjective since we visually inspect the results of each simulation looking for mask violations. Also, each data point comprising the link performance curves is obtained by setting the loss of either the via reflected voltage or the total stripline insertion loss, then sweeping the other variable in 1 dB increments. Therefore the accuracy resolution is expected to be about +/- 0.5 dB; hence, cases can exist that differ slightly but round to be 1 dB apart. Perhaps the largest discrepancy arises where the via reflected voltage is very high. Presumably, experienced SI engineers would not intentionally allow implementation of vias with such high reflected voltages. For example, historically, our guidance for the maximum allowable via reflected voltage has been in the range from -12 to -9 dB.

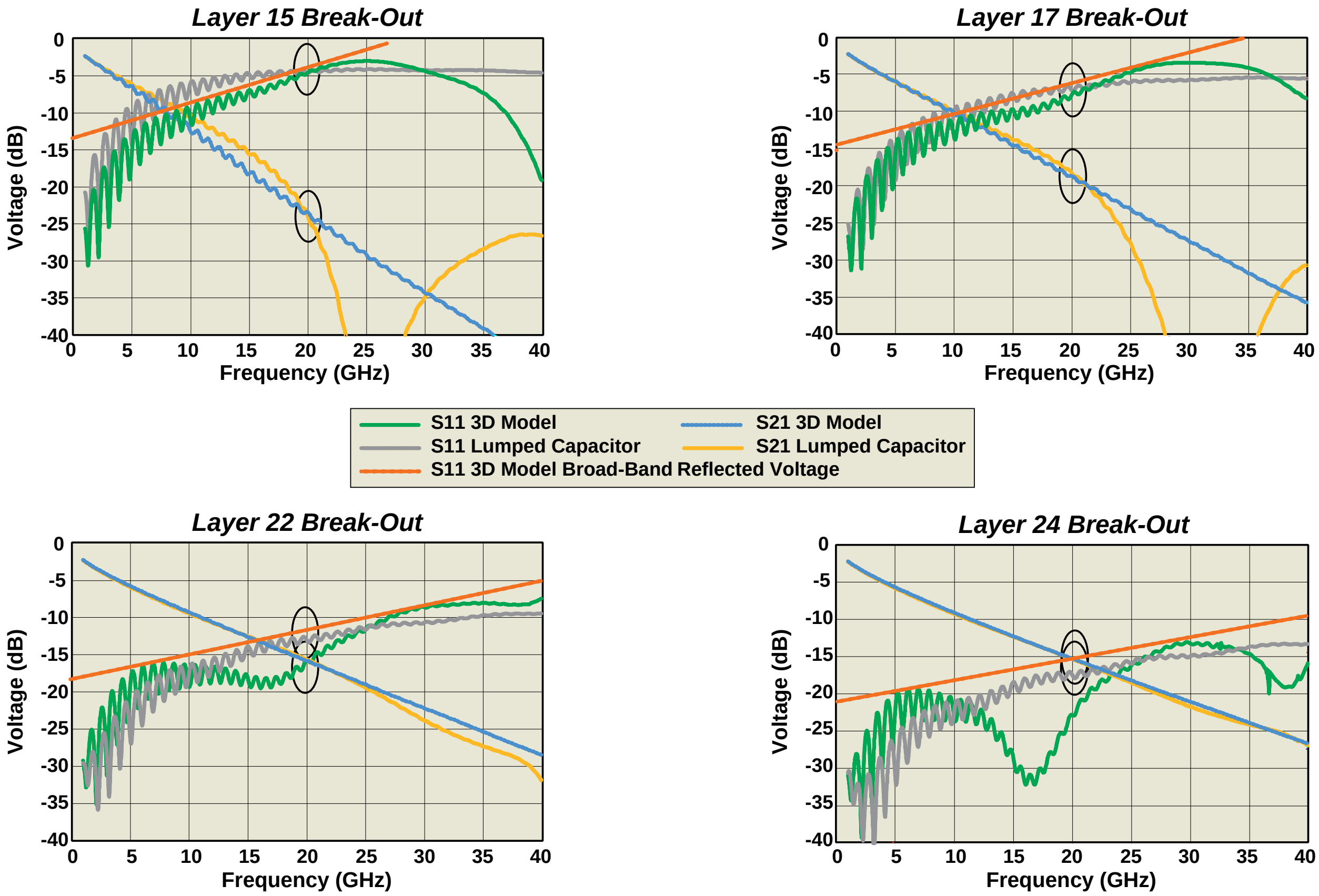

**Figure 6 – Channel S11 and S21 – comparing use of simple capacitor model to 3D model for vias [45444]**

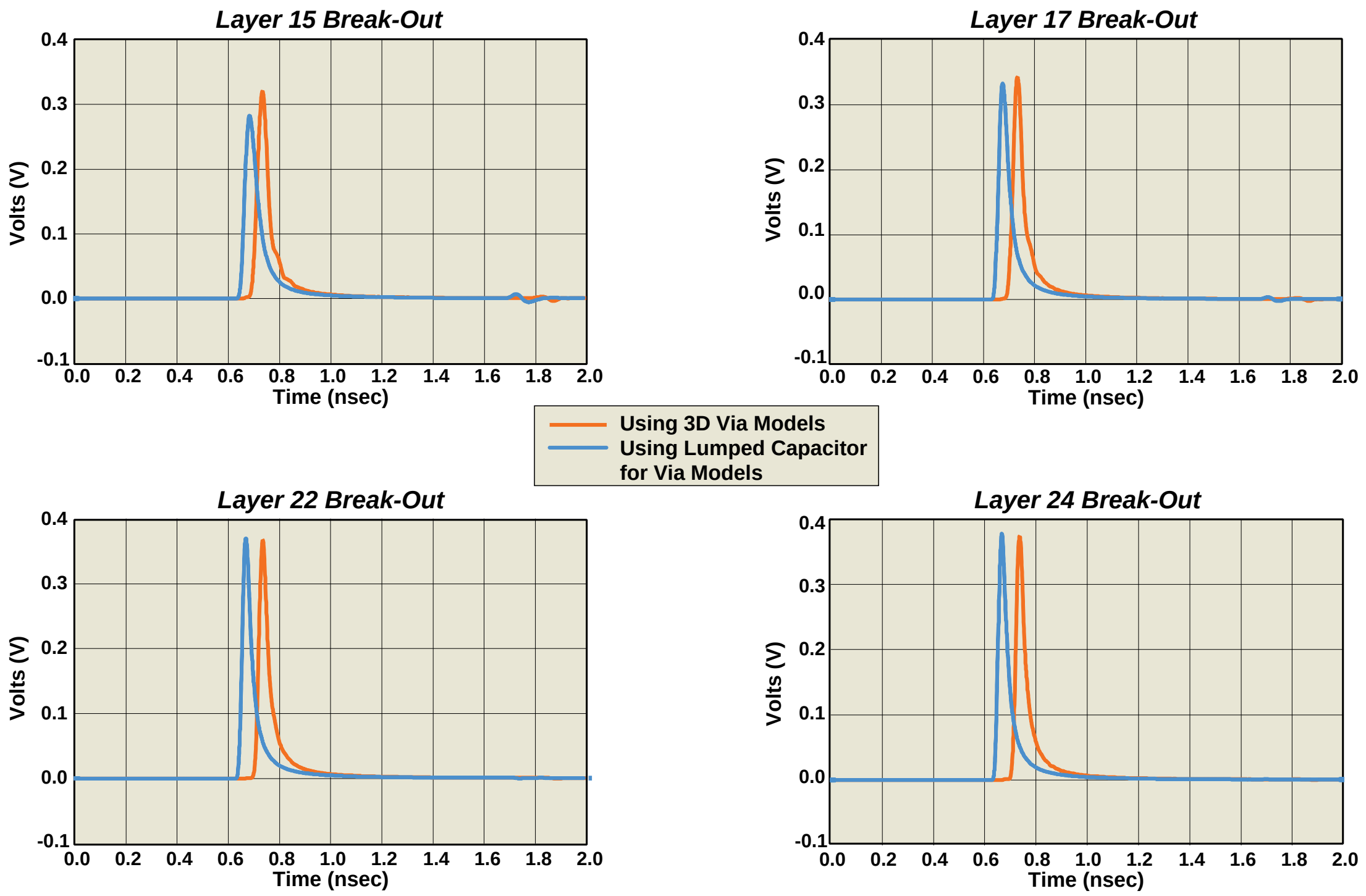

**Figure 7 – 40 Gb/s pulse responses – comparing use of simple capacitor to 3D model for vias [45445]**

## Channel Behavior Studies

The inverted design flow approach, using the simple foundational models, made it very easy to investigate various aspects of channel design. We now discuss four specific case studies. The goal was not to exhaustively evaluate all aspects of passive channel design, but rather to investigate a few specific topics to determine if the results of the link performance matched our expectations. A primary takeaway message is that case studies can be performed efficiently, using this process, as they arise in SI analyses on an actual system. Additionally, many of these case studies can be used to verify accuracy and robustness of the link performance curves.

### Case Study 1 – Via Spacing

This study attempts to determine whether via spacing affects link performance. The DFE is arbitrarily set to use 3 taps. The initial findings indicated that the link performance was essentially unchanged for the three conditions that we studied, with these results plotted in **Figure 8**. We started with a fixed total channel length of 25 UI and placed the vias: 1) in the middle of the channel separated by 2 UI, 2) equally spaced using 8.33 UI of separation, and 3) placed within 1 UI of TX and RX by using a via-to-via spacing of 23 UI. The expected result was that equally spaced vias should offer better performance because vias placed close to one another increase insertion loss ripple by creating large down-and-back reflections between the vias, since there is minimal attenuation along the short stripline segment between the via pair. Similarly, we had expected that vias placed near the TX and/or the RX will cause down-and-back reflections between the via and the IC capacitance. However, in the initial analyses, the IC capacitance and also packaging effects were not included in the channel model, and therefore the reflections were absorbed by the IC terminations. We have noted previously that insertion loss ripple increased substantially when vias were placed closely together. This study suggests that via spacing has little effect on link performance, contrary to the expected results for this simplistic link. Some subsequent modeling did surprisingly show that adding TX and RX capacitances and package models did not appreciably affect this result.

## Case Study 2 – Stripline Loss Mechanisms

For this study we evaluated the sensitivity of link performance to differing types of loss mechanisms, with these results appearing in **Figure 9**. Here, the dependent variable was defined to be the number of DFE taps. It should be noted that all insertion losses were defined at the Nyquist frequency. To simplify this case the vias were removed to better isolate the effects of the various stripline loss conditions. We began by defining a baseline model whereby the loss was split equally across dielectric and metal skin-effect loss mechanisms. Next, when all of the channel loss was assigned to skin-effect, the link performance decreased by approximately 2 dB. However, when all of the loss was assigned to the dielectric, the link performance increased by approximately 5 dB. Therefore, the case with all dielectric loss offers roughly 7 dB better performance than the case with all skin-effect loss. These observations were made assuming 6 DFE taps. Using fewer taps resulted in a smaller performance difference and using more taps produced a larger performance difference. It is unclear why these performance differences were identified. It is possible that the DFE can equalize better for the linear dielectric loss versus frequency compared to the skin-effect loss that is a function of the square root of frequency. Perhaps more likely, due to the frequency dependencies of the two loss mechanisms, the broadband dielectric loss is lower than that of the skin-effect loss and performance is more strongly related to these broadband losses (from DC to Nyquist frequency).

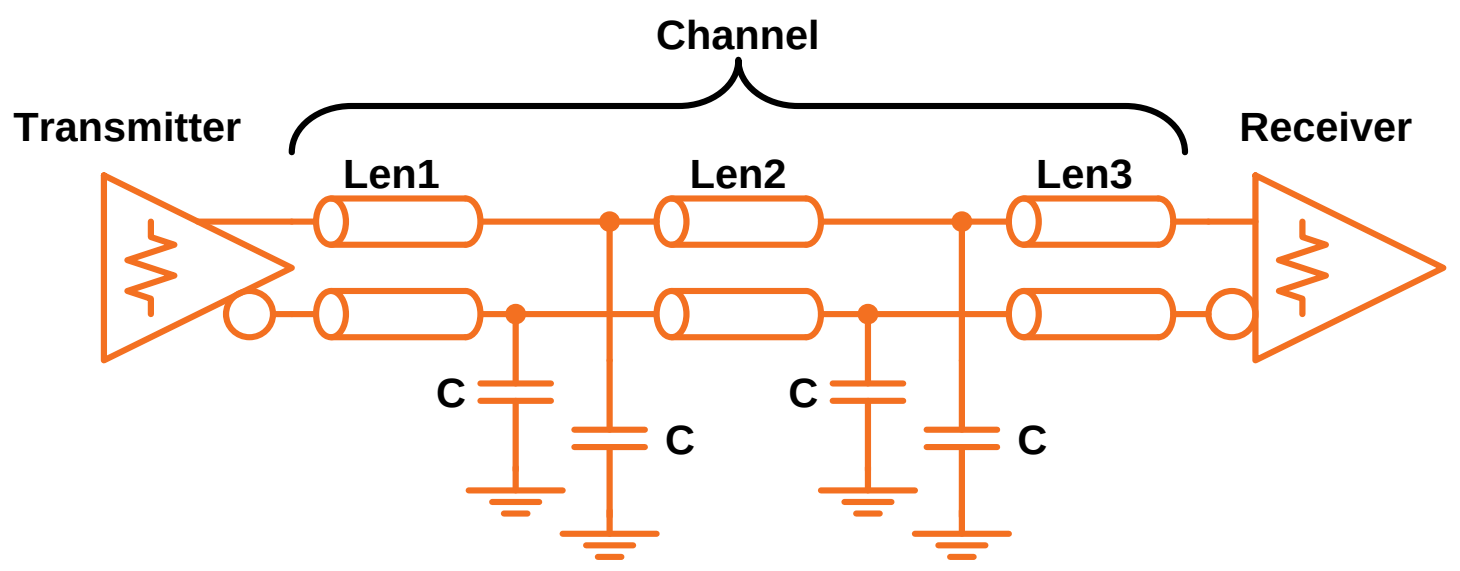

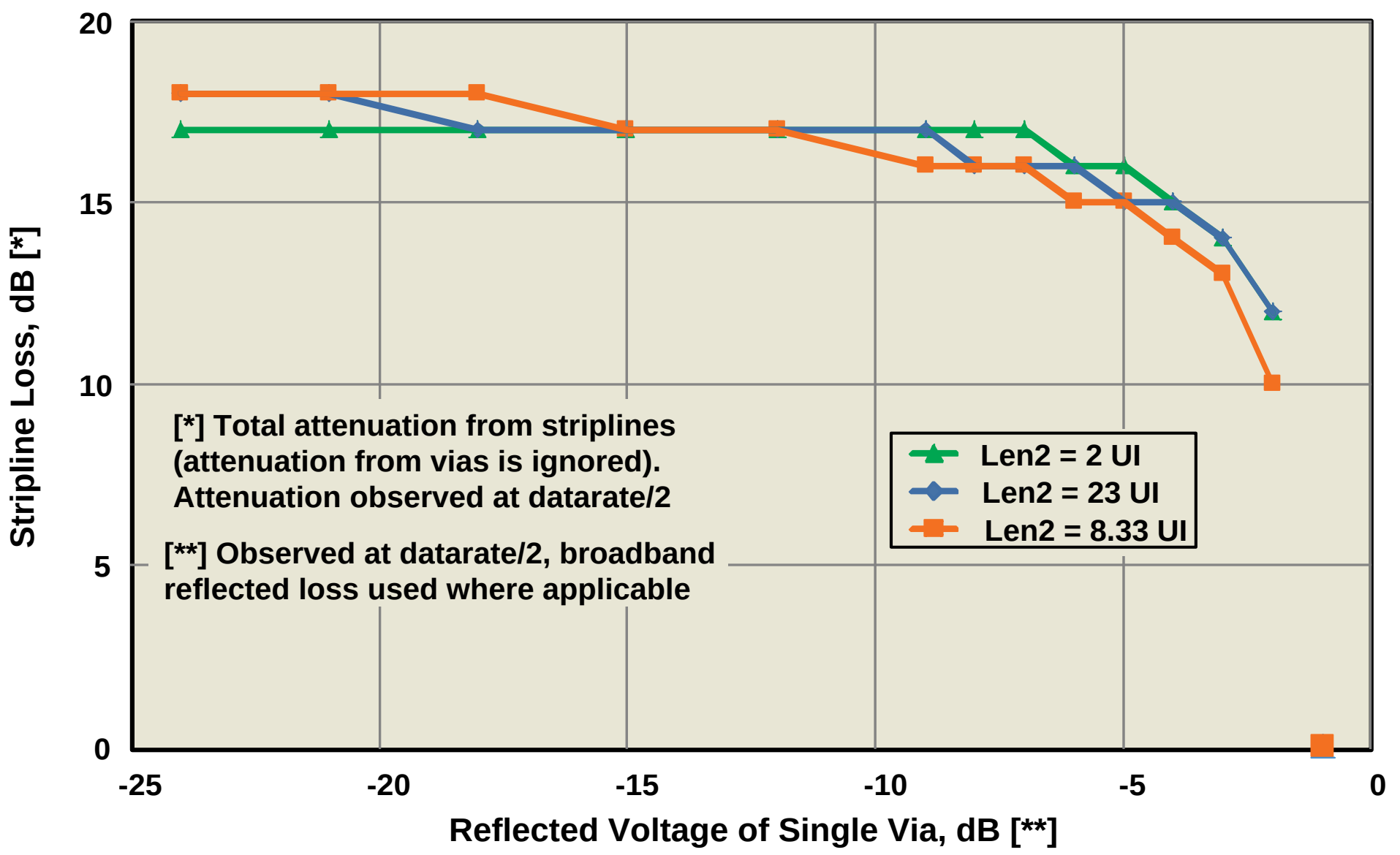

**Figure 8 – Via location effect on link performance [45254v2]**

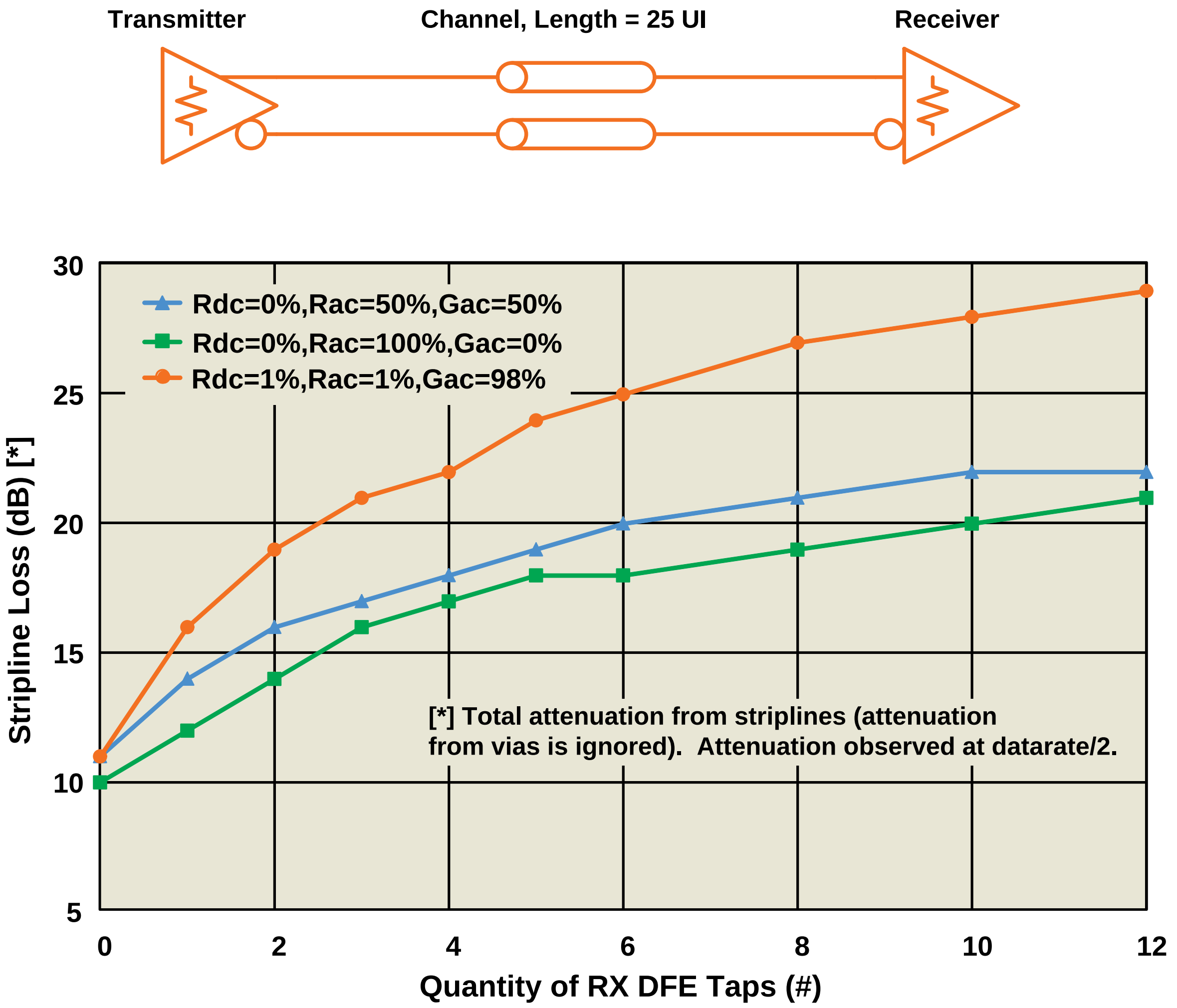

**Figure 9 – Effect that differing stripline loss mechanisms have on link performance [45245]**

## Case Study 3 – Differential Skew

Differential skew was the subject of the third study. For this study, with results presented in **Figure 10**, the DFE was set to use 3 taps and differential channel skew was varied at levels of 0, 0.1, 0.3, and 0.5 UI. The analysis was performed using vias exhibiting reflected voltages of -4 dB (per via), and a second condition where no vias were used in the channel. The results demonstrate roughly 1 and 2 dB drop-off in stripline loss tolerance for 0.3 UI and 0.5 UI of skew, respectively, regardless of whether or not vias were used in the model.

Channel skew of 0.2 UI is offered by **[4]** as a practical limit before link performance becomes noticeably impaired. The results presented here suggest that much higher skew can be tolerated providing that the differential signal lost to common-mode conversion can be offset by, for example, reducing the maximum allowable stripline loss. Such an interpretation requires caution because high levels of skew can create very large common mode voltages that could cause additional RX performance degradation.

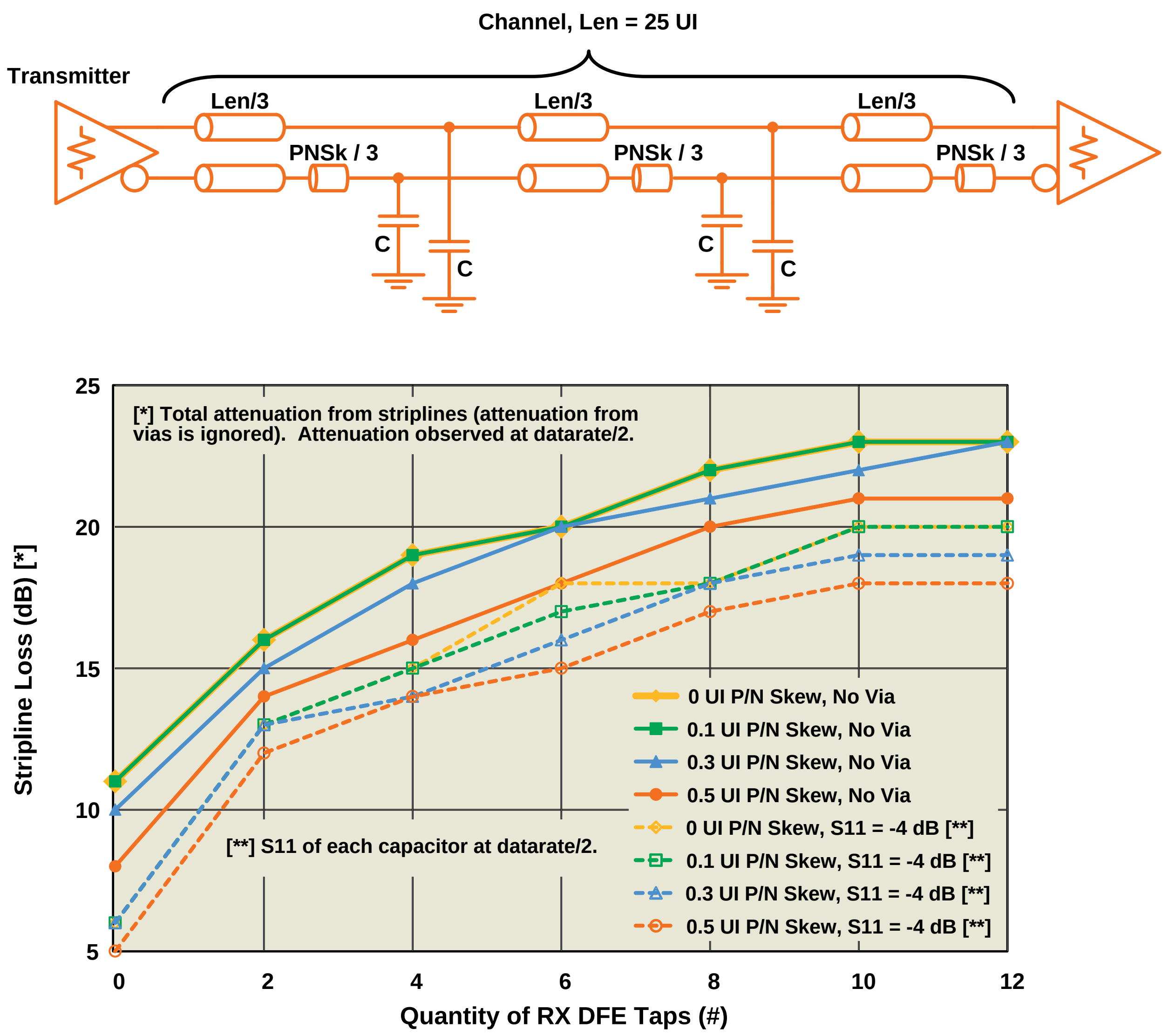

**Figure 10 – Link performance sensitivity to P/N skew [45256]**

## Case Study 4 – Adding Vias Into the Channel

In the fourth and final example, the link performance is documented for the cases in which 2, 3 and 4 vias are added to the channel. In this case the DFE is set to use 3 taps.

In past designs we have had the flexibility simply to add routing layers into the PCB stack-up to avoid adding layer-transition vias for crossing over blocked channels. The recommendation to minimize usage of layer-transition vias has been adapted widely within the industry **[5] [6] [7]**. Using the SerDes link design flow process that we described previously, it is fairly easy to determine, with a high degree of confidence, whether we can add layer-transition vias to bypass routing blockages. For example, note in **Figure 11** that if the via reflected voltage can be held below -8 dB, then the performance does not change significantly when using 2 versus 4 via pairs.

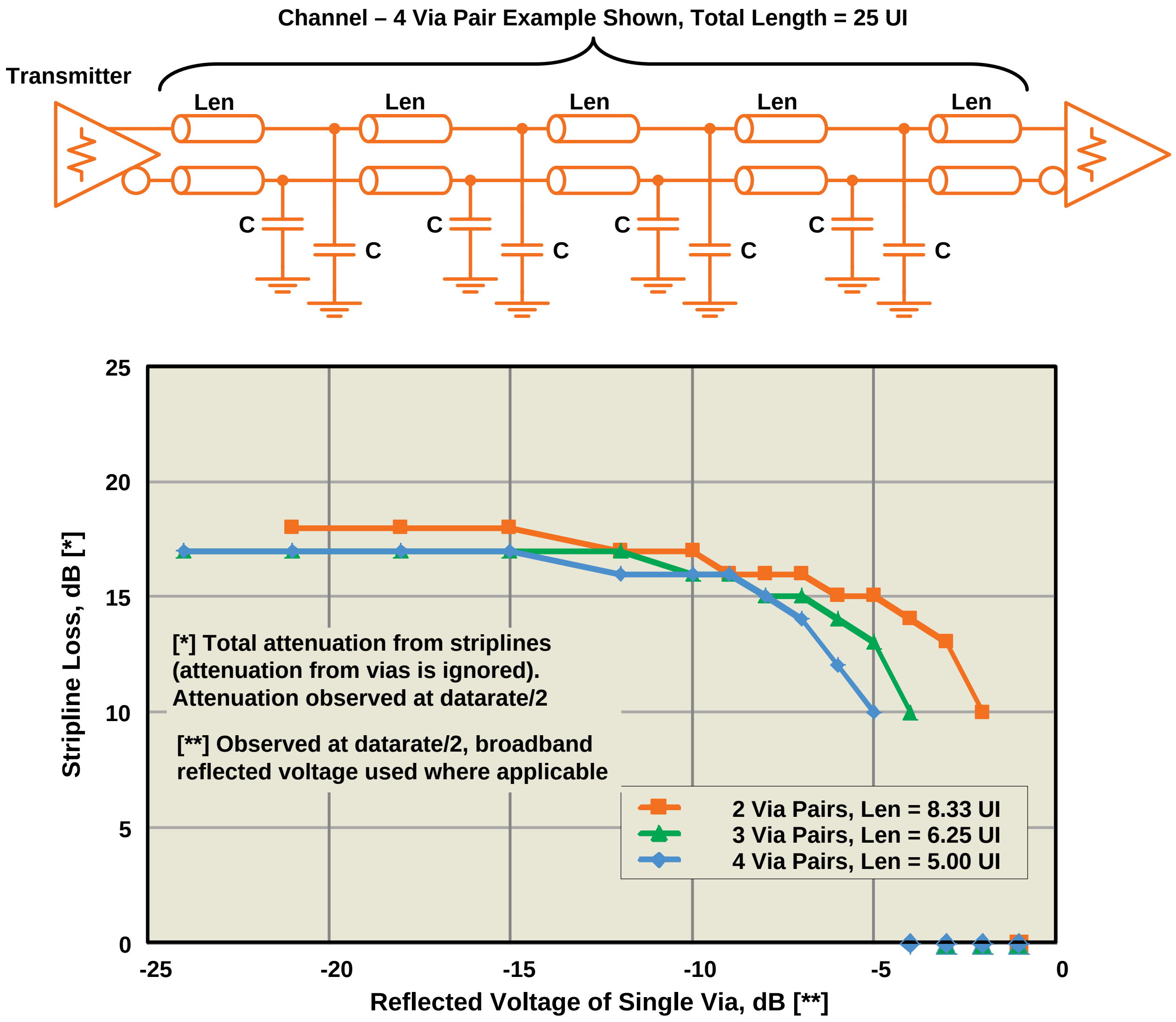

**Figure 11 – Link performance – two, three, and four vias [45463]**

## Incorporating Crosstalk Into Link Performance Modeling

Crosstalk will be more of a concern as the speed and density of SerDes links increase. Foundational channel models proved to be very useful in the creation of link performance curves, as stated. However, these models did not address crosstalk, which does need to be accounted for in some manner in order to obtain an accurate estimation of link performance.

For simplicity, in this work it has been assumed that a large majority of crosstalk originates from coupled vias and that the PCB and package designs are such that stripline crosstalk is negligible. Opposing channels, as illustrated in **Figure 12**-a, directly inject near-end crosstalk (NEXT) into a victim channel. Channels with the same flow, as shown in **Figure 12**-b, directly inject far-end crosstalk (FEXT) into a victim channel. Some of the crosstalk not injected directly towards the RX can be reflected from vias and other channel impairments back toward the RX, and therefore one should generally account for both NEXT and FEXT.

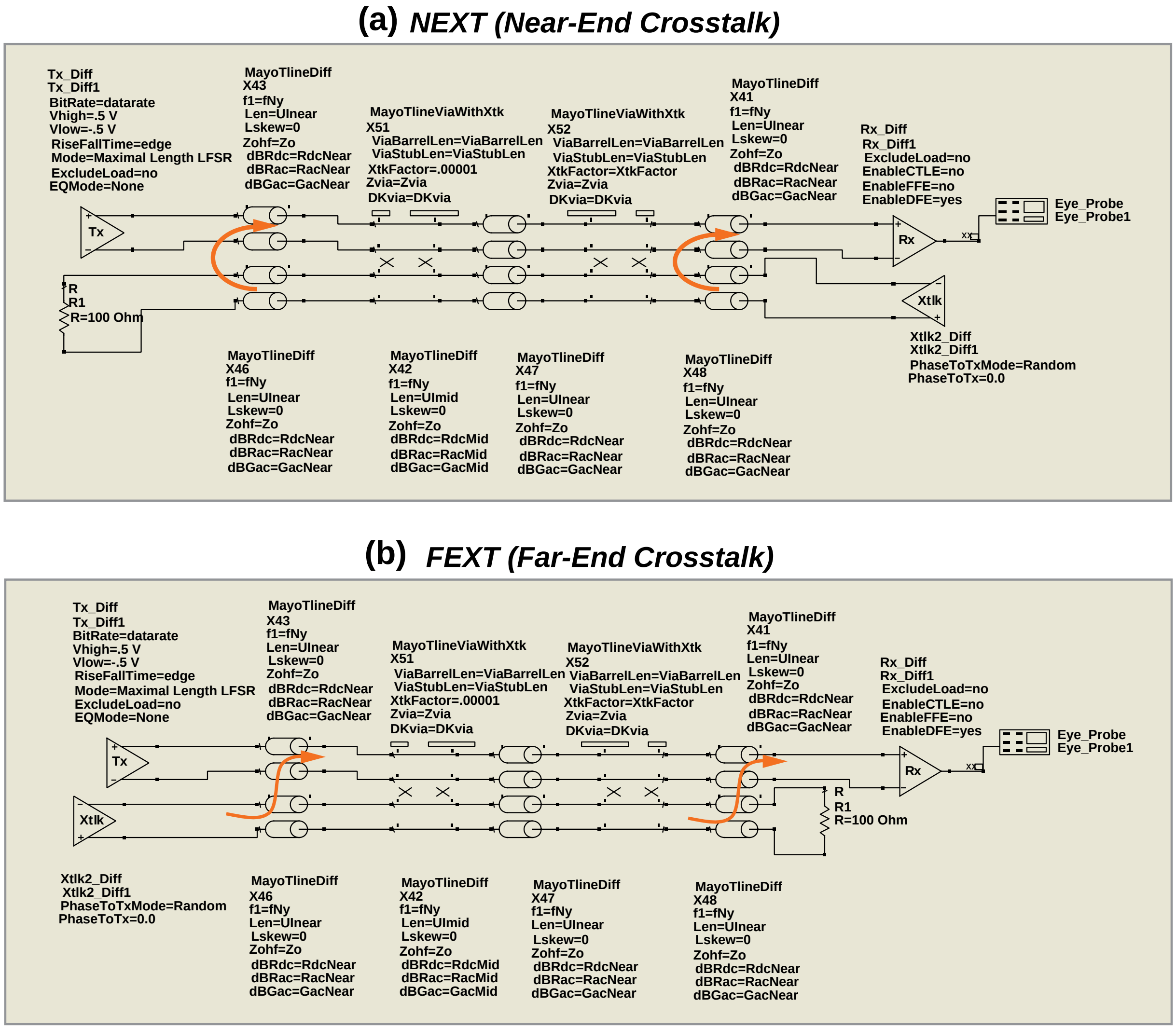

**Figure 12 – Channel crosstalk – near and far-end examples [45447]**

Victim and aggressor channels typically do not couple together along the full length of the channel. Instead, an aggressor channel may inject noise near the TX and a different aggressor may inject noise near the RX with these respective aggressor channels otherwise being uncoupled from the victim channel. Moreover, each grouping of vias, such as a pin-field break-out beneath a ball-grid-array (BGA) package, could have several aggressors injecting crosstalk onto a victim. We have not yet begun aggregating all sources of crosstalk, but to date have studied the model accuracy of crosstalk injection within a coupled via structure. Analyzing crosstalk is a complicated endeavor and perhaps a foundational model would minimize the level of effort required for a thorough crosstalk analysis. Therefore, in this section, we investigate methods to quickly model crosstalk to help develop a broad understanding of crosstalk behavior.

Perhaps the simplest foundational model to consider that incorporates crosstalk is realized by extending the lumped-capacitor foundational model for the via by adding a cross-coupled capacitor across the victim and aggressor nodes, as depicted in **Figure 13**-a. Earlier it was stated that good matching to either time and/or frequency "known-good" responses is not a firm requirement for a foundational model, but rather that it was singularly important that link performance curves be accurate when using foundational models. Those assessments were based on observations using an isolated channel model. Applying this principle to crosstalk-coupled channels, the most relevant metric is that using the foundational crosstalk models will degrade the link performance curves similarly to those when using the 3D models. However, link performance curves using the coupled capacitor models did not match well to those using the 3D model. Therefore, a different approach to create an improved foundational model that better represents via crosstalk was needed.

From our prior experience, via crosstalk behaves in a similar manner to inductive coupling. This observation is reasonable since the PCB power and ground planes capture electric field between victim and aggressor vias and therefore greatly reduce capacitive coupling between these vias. However, these planes do not provide a continuous path for return currents and therefore these planes provide very minimal inductive shielding between victim and aggressor vias. Thus we attempted to use inductive coupling between the victim and aggressors. Inductively coupled circuits require (self) inductors; hence we changed the shunt capacitive circuits to series inductor circuits using a mutual coupling factor between the aggressor and victim inductors as presented in **Figure 13**-b. Reflected voltages in the time domain are the same magnitude but of opposite polarity for series-inductive and shunt-capacitive circuits. Likewise, frequency responses for both circuits have the same magnitude response but their phase responses differ by 180 degrees. Therefore, ignoring crosstalk, it was unsurprising that we found the link performance curves to match closely when using inductive or capacitive via models. These observations led us to experiment with coupled inductor models for the vias, which might offer more accurate crosstalk modeling accuracy even though the typically capacitive vias are modeled as being inductive.

Both capacitive and inductive models have identical crosstalk magnitude responses versus frequency as in **Figure 13**-c, where we set the crosstalk factor for the custom models to be -46 dB at 20 GHz. Unfortunately, for high values of reflected voltage, the crosstalk is lower than specified by the equations in the model, because our equations assume that the reactance of the via model has no effect on the crosstalk. Therefore, we note a several-dB reduction in specified crosstalk when the reflected voltage rises from -15 to -2 dB.

We also found the inductive models of the coupled vias to be inaccurate in accounting for the effects of crosstalk in the link performance curves. This result was not surprising since similar equations were used to specify crosstalk whether using the capacitive or inductive via models.

However, the crosstalk response polarity shifts from being in-phase with a positive NEXT and FEXT (**Figure 13** d/e) using the capacitive circuit to out-of-phase with negative NEXT and positive FEXT (**Figure 13** f/g) using the inductive circuit. This result suggested the use of an inductively-coupled transmission line as a means to create a foundational coupled via model generating accurate crosstalk effects.

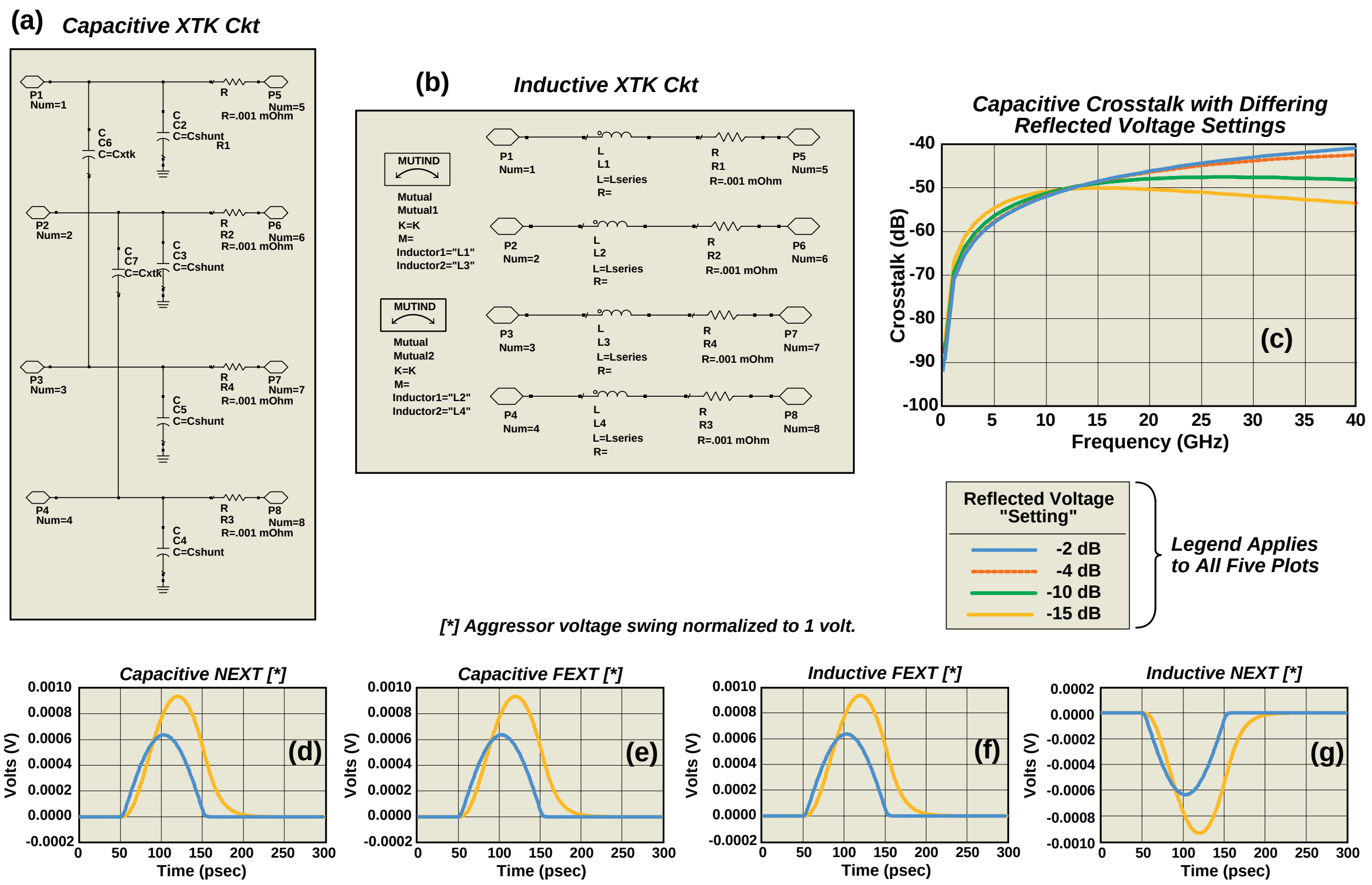

**Figure 13 – Crosstalk behavior of simple inductive and capacitive circuits [45448]**

We investigated a quasi-physical model consisting of a transmission line to represent the via barrel from the top of the PCB to the stripline egress, with the addition of another short transmission line to emulate the behavior of the via barrel stub. Such a model appears in **Figure 14**. Inputs to this model are the via barrel and via stub lengths, the differential characteristic impedance of the via, the dielectric of the PCB material surrounding the via, and the crosstalk factor. Presently the crosstalk factor is manually adjusted until the simulated crosstalk matches well to that of a 3D model. Eventually, we hope to implement an explicit definition for the crosstalk rather than an arbitrary factor that requires adjustment, but first the viability of this model must be addressed.

The equations in **Figure 14** convert the user-supplied parameters to definitions for the ADS CLINP coupled transmission line model. One coupled pair represents the victim and aggressor true side of a differential pair. The other coupled pair represents the victim and aggressor complement side of a differential pair. The true/complement pairs are not actually coupled even though the coupling strength of differential vias is typically 20 dB stronger than that of the crosstalk coupling magnitude. Ignoring the differential pair coupling simplifies the model and should not affect overall modeling accuracy provided that well balanced signals are used.

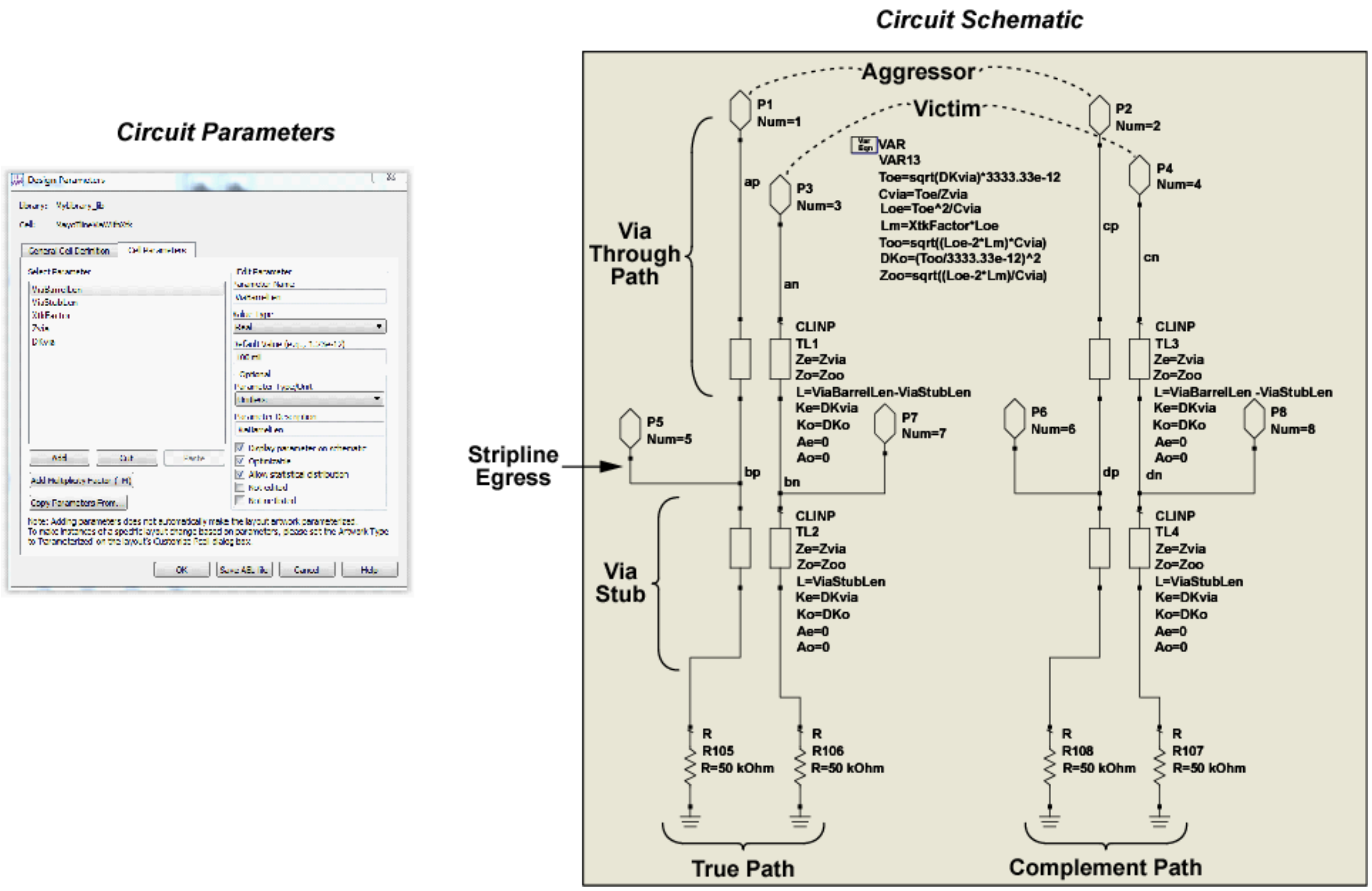

**Figure 14 – Coupled transmission line (t-line) used to model crosstalk in differential via pair [45461]**

Referring to **Figure 15**, the reflected voltage of the "t-line" via models compares well to that of the 3D models for the cases with longer via stubs, but less favorably for shorter via stubs. We believe that this discrepancy is related to the effects of real 3D structures not represented in the simplified t-line via model, such as the inductive behavior of striplines passing over the antipad voids **[3]** to connect to the via barrel, and also due to the terminating pads on the bottom of the via barrels. We can eliminate these stripline and pad structures in a via model that has only top and bottom-side ports, and model this 3D structure to achieve results (not shown) that match very well with the t-line via model.

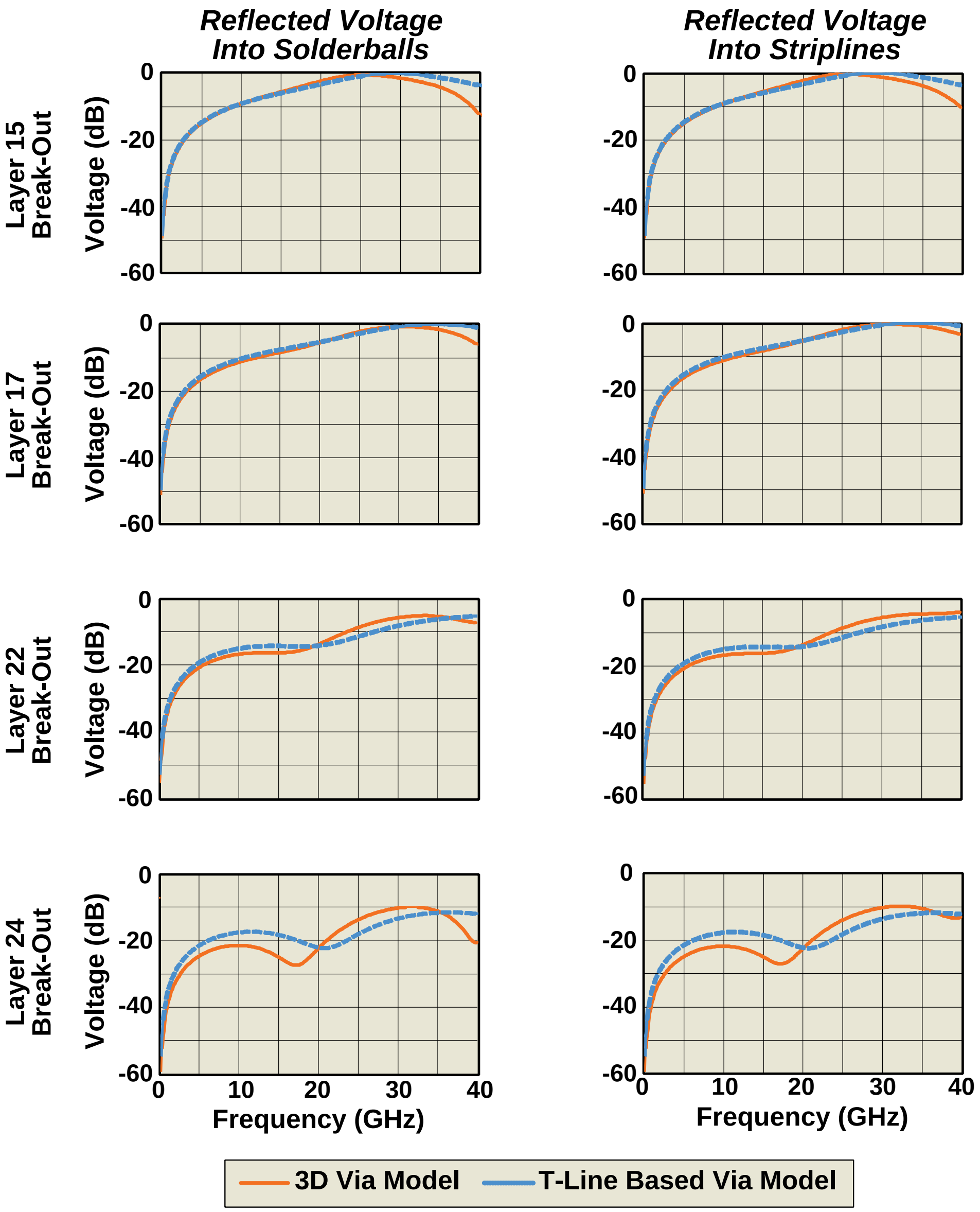

**Figure 15 – Via model reflected voltage (S11) comparisons - coupled t-line versus 3D crosstalk models [45472]**

Referring to **Figure 16**, the frequency-dependent crosstalk from the t-line and 3D via models match nearly perfectly for FEXT, but the matching is not as favorable or consistent for the NEXT across the four different models. FEXT, in either the time or frequency domain, is not dependent on the direction that the aggressor signal is driven, but we include all results for completeness.

Using the simulated S-parameters from the 3D models and the t-line via models we plotted the crosstalk in response to a voltage step, as charted in **Figure 17**. The NEXT match (first column) between the results using the t-line via model is very close to those using the 3D model if the aggressor pulse is driven from the solder ball side, i.e., from the top of the PCB surface. However, the NEXT match (third column) is quite poor with regard to pulse magnitudes if the aggressor step voltage is driven from the striplines. The FEXT crosstalk responses using the t-line model are reasonably close to that using the 3D model.

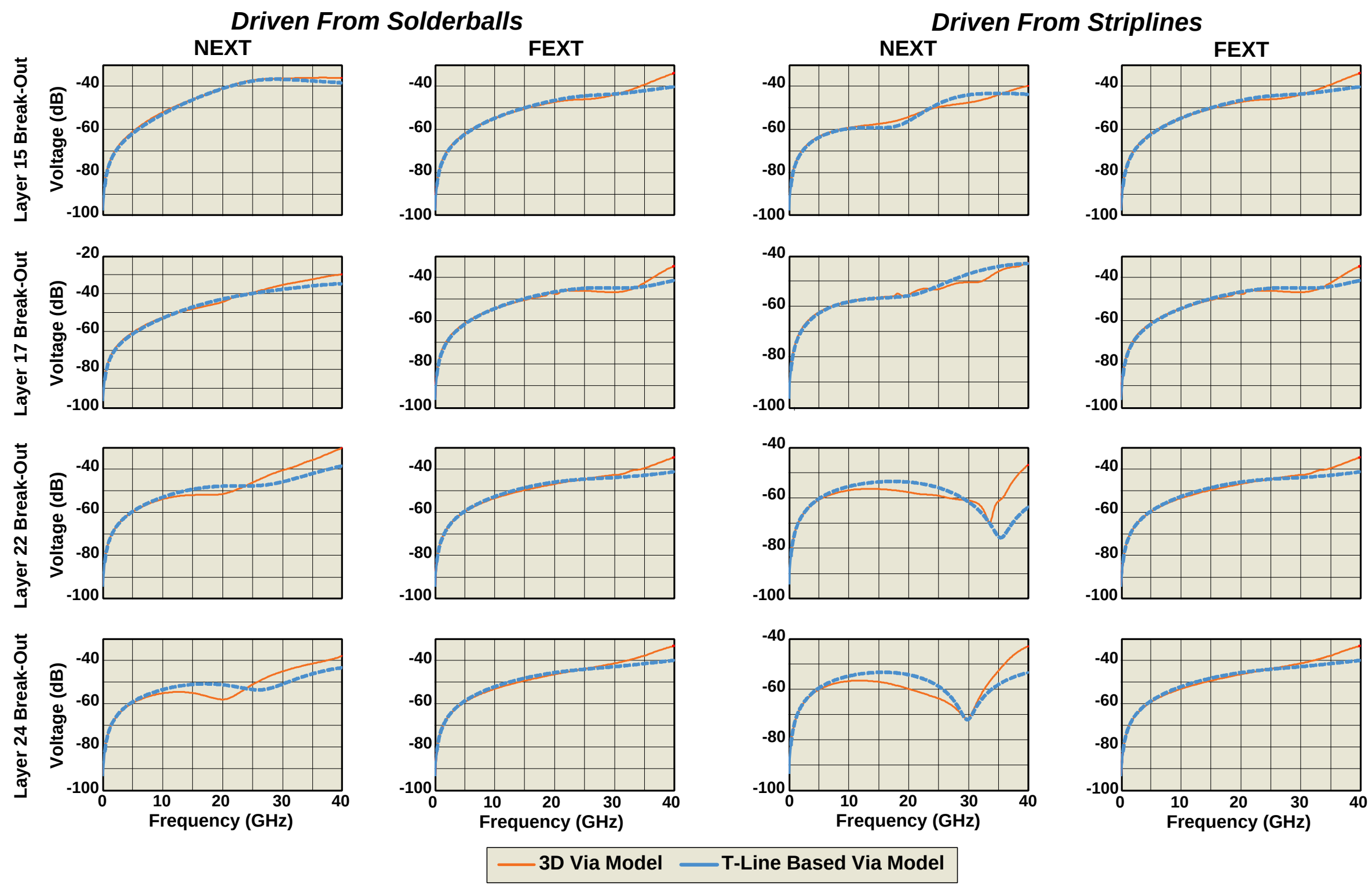

**Figure 16 – Via model frequency domain comparisons - coupled t-line versus 3D crosstalk models [45458]**

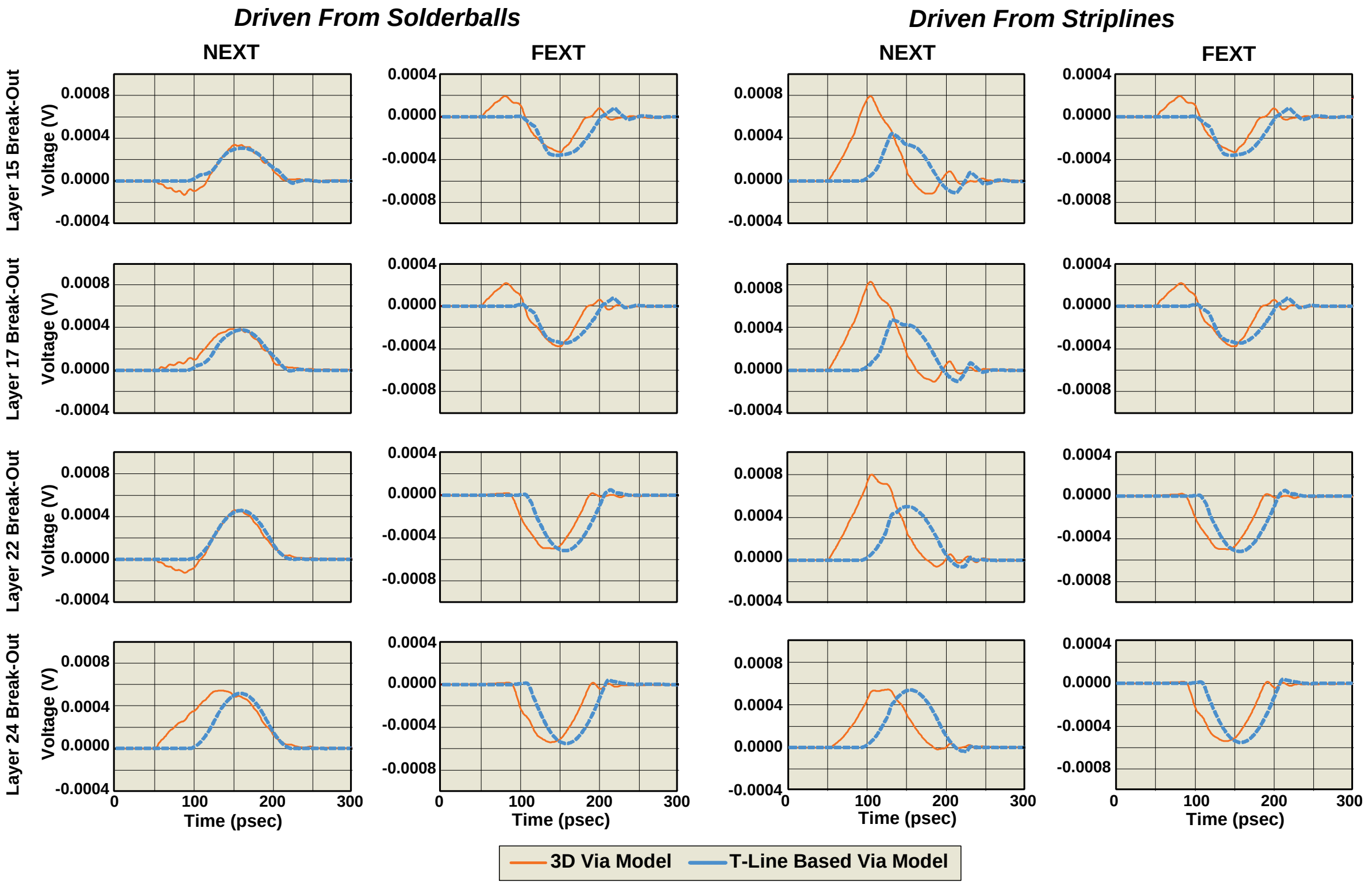

**Figure 17 – Via model time domain comparisons - coupled t-line versus 3D crosstalk models [45459]**

Providing that a simplistic via model with accurate crosstalk can be developed, it may be useful to determine how this model might be used in order to assess the impact on link performance. Does it make sense to try to simulate directly the crosstalk affecting the link performance by adding this via model into the channel model? How might crosstalk modeling be implemented for complex systems having hundreds of links each with many crosstalk aggressors? Could there be a methodology to determine the impact of crosstalk on eye closure such that crosstalk need not be directly included in the link simulations (such as with root mean square (RMS) summation methods used for some existing link standards)? We are still investigating the best methodology to accommodate the effect of crosstalk on SerDes links. However, we undertook an initial analysis that may be useable to extrapolate crosstalk effects on link performance using frequency domain crosstalk results.

To understand crosstalk behavior more thoroughly, the results in **Figure 18** are presented to show the effectiveness of translating frequency-domain crosstalk to time-domain crosstalk from a pulse response. This study was conducted at a Nyquist frequency of 10 GHz (for frequency-domain data on the left-hand side) and associated data rate of 20 Gb/s (for time-domain data on the right-hand side of the figure). Frequency data is presented in millivolts rather than in dB to allow for a direct comparison to time-domain crosstalk peak-to-peak voltages. All of these results were based on the t-line via model which should be very well-behaved in both frequency and transient analysis, whereas 3D simulations have historically had accuracy issues in transient simulations. The t-line via model allowed us to sweep the via stub length, which is the dependent axis for these plots.

One of the problems with obtaining frequency-domain crosstalk is that the crosstalk varies over frequency – often presenting a “valley” at the desired measurement frequency point. Therefore we alternatively offer a “broadband” value where the crosstalk is taken at the Nyquist frequency along a straight line drawn across the “humps” of the frequency-dependent crosstalk. This approach is analogous to the BRV method previously described. From these results, in general we note that the frequency-based crosstalk translates very well to that observed using a pulse response, although the FEXT driving stripline case, for longer stub lengths, somewhat underestimates the crosstalk observed in the transient simulation.

The results show some promise in extrapolating frequency-domain crosstalk to time-domain crosstalk, at least for the case in which the time-domain crosstalk is measured in terms of peak-to-peak values in response to a pulse stimulus. Ultimately our goal is to extrapolate frequency-domain crosstalk to eye closure or some other link performance metric. An initial attempt to model eye closure when the crosstalk model was added into the channel was not successful, since it was difficult to measure additional eye closure of only a few millivolts. It is possible, for example, to increase the crosstalk coupling by an order of magnitude in an attempt to measure the eye closure accurately. Since eye closure due to crosstalk is a broadband function related to the bit-stream pattern, it will be of interest if a simple and accurate methodology can predict eye closure from crosstalk in the frequency-domain.

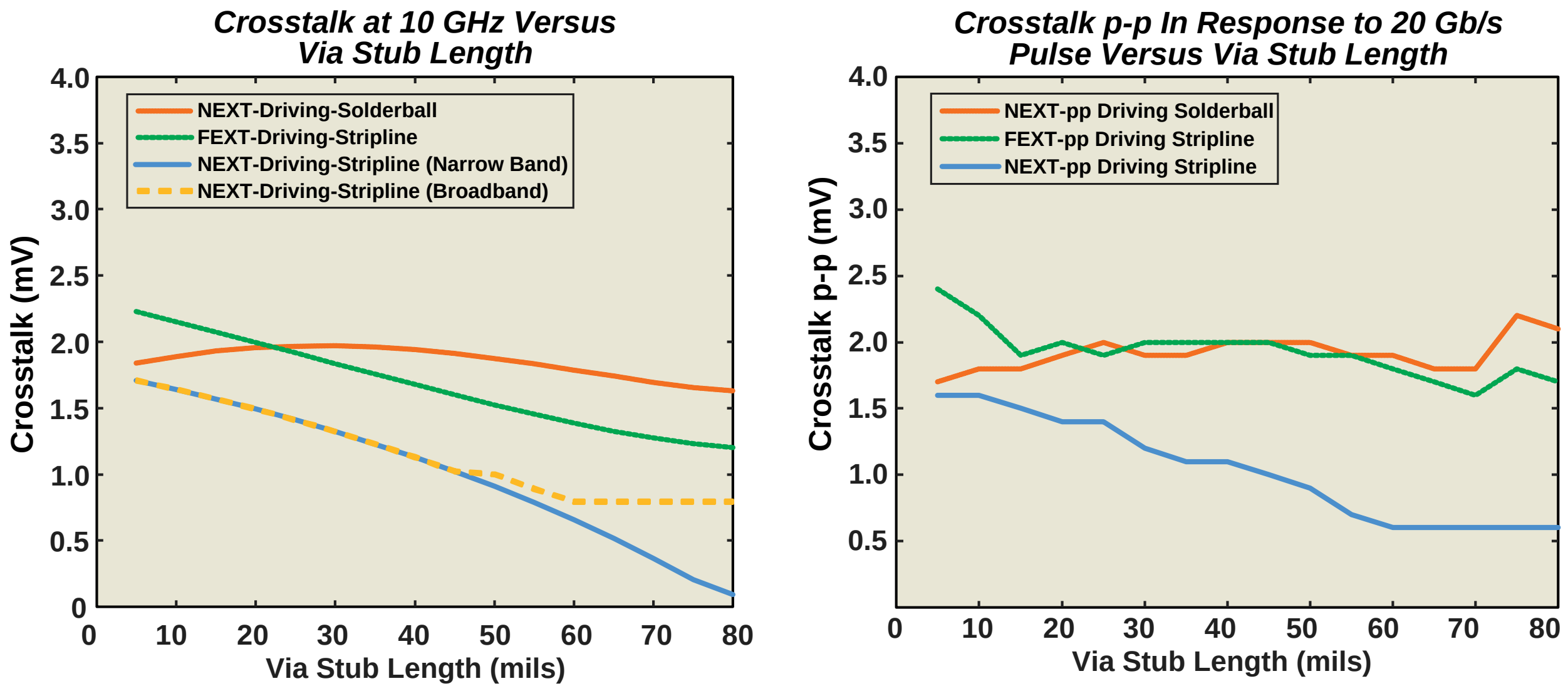

**Figure 18 – Crosstalk of coupled t-line via model – frequency versus time domain [45460]**

# Summary

We inverted the traditional SerDes link signal integrity analysis process as a means to meet future simulation and modeling requirements for systems having increased data rates and larger numbers of SerDes links. To facilitate the generation of link performance curves we modeled the SerDes channels using foundational models, which are defined by fundamental electrical properties, in contradistinction to the complex and detailed 3D electromagnetic models used traditionally.

We have demonstrated that, using these foundational models, link performance can be modeled accurately for the simplified links described in this paper, despite the fact that the foundational models did not match well to either time-based or frequency-dependent responses.

The foundational models, due to their simplicity, were useful in a variety of link behavior studies. We have described four examples of such studies, including cases that yielded results contrary to expectations.

The inverted design flow process did not accurately account for crosstalk when using the simplest foundational models. Better results were achieved with a mixed foundational/physical via model; this work is still in progress. It is still not clear whether to include crosstalk models directly in the link simulations. Alternatively, techniques such as RMS summation of the crosstalk contributors can be employed to determine crosstalk levels for particular channels. It would then be possible to budget for this crosstalk, e.g., by growing an eye mask by the amount of the expected crosstalk. Using this approach it would not be necessary to model crosstalk during full link simulations and the simplest foundational models for PCB vias could be used.

A subject not discussed above was the methodology to identify the best TX and RX equalization and related settings. In the past we employed several different methodologies to optimize equalization settings, depending on the capabilities of the SerDes technologies and the sophistication of the vendor-supplied TX and RX models. In the future the efficiency of the inverted link modeling process may be used to achieve a better understanding of optimizing equalization settings.

Since the area of concentration of this paper has been on the link analysis process, readers are cautioned against using the specific results directly since the channel models were incomplete, with limited TX and RX equalization, in an attempt to present the modeling process more clearly. We would recommend modeling the links using essentially complete channels by adding IC capacitances, first-level package models, the expected number of vias, and any connectors used in the system.

Finally, our primary recommendation is to begin a system SerDes link design process by first creating the link performance curves and using these curves to choose PCB technologies, channel topologies, and so on. Point studies should be undertaken using the foundation models for expediency to validate these link performance curves. We now have a good understanding of how foundational stripline models can be realized in differing PCB technologies, and can also readily equate the reflected voltage figure of merit for foundational via models to physical PCB via designs. We have introduced the use of foundational models to expedite the creation of link performance curves and to conduct point studies. However, physically-based models can alternatively be used for the inverted link design flow with varying degrees of success and increased effort.